\documentclass{aa}  
\usepackage{graphicx}
\usepackage{txfonts}
\usepackage{xcolor}
\usepackage{amsmath}
\usepackage[breaklinks, colorlinks, citecolor=blue, linkcolor=blue]{hyperref}
\DeclareMathOperator*{\argmax}{arg\,max}
\DeclareMathOperator*{\argmin}{arg\,min}

\newcommand{\alphabold}{\mbox{\boldmath$\alpha$}}

\begin{document} 
  
   \title{Learning to do multiframe wavefront sensing unsupervisedly: applications to blind deconvolution.}
   % \title{Unsupervised learning to blindly estimate the point spread function}

   \author{A. Asensio Ramos\inst{1,2}, N. Olspert\inst{3}}%, M. van Noort\inst{3}}
   % \author{A. Asensio Ramos, A. Oscoz, P\'erez-Prieto, J. A., L\'opez, R.}

   \institute{Instituto de Astrof\'{\i}sica de Canarias, 38205, La Laguna, Tenerife, Spain; \email{aasensio@iac.es}
\and
Departamento de Astrof\'{\i}sica, Universidad de La Laguna, E-38205 La Laguna, Tenerife, Spain
\and
Max Planck Institute for Solar System Research, Justus-von-Liebig-Weg 3, 37077 G\"ottingen, Germany}

   \date{}

 \abstract
 {Observations from ground based telescopes are affected by the presence of the 
 Earth atmosphere, which severely perturbs them. The use of adaptive optics techniques
 has allowed us to partly beat this limitation. However, image selection or
 post-facto image reconstruction methods applied to bursts of short-exposure images
 are routinely needed to reach the diffraction limit. Deep learning has
 been recently proposed as an efficient way to accelerate these image reconstructions.
 Currently, these deep neural networks are trained with supervision, so that either standard
 deconvolution algorithms need to be applied a-priori or complex simulations of the
 solar magneto-convection need to be carried out to generate the training sets.}
 {Our aim here is to propose a general unsupervised training scheme that allows multiframe
 blind deconvolution deep learning systems to be trained simply with observations. The approach 
 can be applied for the correction of point-like as well as extended objects.}
 {Leveraging the linear image formation theory and a probabilistic approach to the blind
 deconvolution problem produces a physically-motivated loss function. The optimization of
 this loss function allows an end-to-end training of a machine learning model composed of three neural 
 networks.}
 {As examples, we apply this procedure to the deconvolution of stellar data from the 
 FastCam instrument and to solar extended data from the Swedish Solar Telescope.
 The analysis demonstrates that the proposed neural model can be successfully trained without supervision
 using observations only. It provides estimations of the 
 instantaneous wavefronts, from which a corrected image can be found using 
 standard deconvolution technniques. The network model 
 is roughly three orders of magnitude faster than applying standard deconvolution based 
 on optimization and shows potential to be used on real-time at the telescope.}
 {} 
% 5 {} token are mandatory
 
 % \abstract{.}
  % {bla}
  % {The stability equations of state are
  % calculated for solar composition and are displayed in the domain
  % $-14 \leq \lg \rho / \mathrm{[g\, cm^{-3}]} \leq 0 $,
  % $ 8.8 \leq \lg e / \mathrm{[erg\, g^{-1}]} \leq 17.7$. These displays
  % may be
  % used to determine the one-zone stability of layers in stellar
  % or planetary structure models by directly reading off the value of
  % the stability equations for the thermodynamic state of these layers,
  % specified
  % by state quantities as density $\rho$, temperature $T$ or
  % specific internal energy $e$.
  % Regions of instability in the $(\rho,e)$-plane are described
  % and related to the underlying microphysical processes.}
  % {Vibrational instability is found to be a common phenomenon
  % at temperatures lower than the second He ionisation
  % zone. The $\kappa$-mechanism is widespread under `cool'
  % conditions.}

   \authorrunning{Asensio Ramos et al.}
   \titlerunning{Learning to do multiframe blind deconvolution unsupervisedly}
   \maketitle
%
%________________________________________________________________

\section{Introduction}
The observation of astronomical objects from ground-based observatories
is degraded by the presence of turbulence on the Earth atmosphere. One obvious solution
is to move the observatory to space to avoid the atmosphere, but this is often not feasible
due to technological or budgetary reasons. Additionally, the largest
and more advanced telescopes are always built on the ground, because they
usually need technology at the forefront of science.

Active and, especially, adaptive optics (AO), i.e., deformable optics that can compensate for
the effect of the atmosphere almost in real time, have facilitated
the use of ground based telescopes. The combination of very fast sensors (that
allow the measurement of instantaneous wavefront) and deformable mirrors (that correct
the wavefront that reach the science cameras) can produce images
very close to the diffraction limit of the telescopes, at least in a reduced
field-of-view (FOV). As a demonstration of the importance of AO, currently more 
than 25\% of the observations in some large aperture telescopes like Keck and VLT 
use any AO device \citep{rigaut15}. This figure goes up to $\sim$100\% in the
case of solar observations.
These AO systems have been really succesful in the near
infrared, where the perturbing effect of the atmosphere is less important.
AO systems working for visible and near ultraviolet wavelengths are
much more demanding and, although lagging behind, many efforts
are put on making them robust.
A case of success of using AO systems in the visible is the field of solar physics, where such
systems are commonly used in telescopes like the Swedish 1-m Solar Telescope
(SST) at the Observatorio del Roque de los Muchachos (Spain), the GREGOR
telescope on the Observatorio del Teide (Spain) or the Goode Solar Telecope
(GST) on the Big Bear Observatory (USA).

Even if AO systems are working properly, some residual wavefront perturbations are
still present on the images. These residuals are a consequence of the
accumulation of different sources: i) the
wavefront sensors (WFS) are not measuring the wavefront perfectly, ii) the deformable
mirrors are not correcting properly the wavefront measured by the WFS, iii) there 
is some delay between the measurement and the actuation, iv)
static aberrations in the telescope+instrument optics are not corrected by AO
systems and v) classical AO systems with one WFS and one pupil deformable mirror produce
their best correction close to the optical axis, so the rest of FOV has a much worse correction.

For the previous reasons, reaching the diffraction limit of a telescope in a large FOV is not often
possible without a posteriori image correction methods. The simplest techniques
of a posteriori correction are based on frame selection, also known as
\emph{lucky imaging}. These methods
rely on the fact that the wavefront deformation due to the atmosphere is
small at some selected frames when a long burst of short-exposure images is
acquired. The fraction of such \emph{lucky} frames decreases when the 
atmospheric turbulence increases. Another problem with this technique lies in 
its low photon efficiency, because a very large fraction of the frames are discarded. 
An additional drawback is that it only works properly for small or medium-sized
telescopes, with diameters below 2.5 m. In larger
telescopes, the probability that low turbulence is found in a significant
fraction of the telescope aperture quickly goes to zero.
Instruments like FastCam \citep{oscoz08}, that we use in this paper, 
are fully based on the exploitation of this idea.

More elaborate techniques are based on speckle methods \citep{labeyrie70,vonderluhe93},
which make use of all the recorded frames to get an estimation of the image. 
\cite{paxman92} later proposed some improvements, which are now at the
base of many of the most advanced methods currently in use. The first one was the 
assumption of a very flexible parametric point spread function (PSF),
that is specially suited for telescopic observations. This is done via a
linear expansion of the aberrations in the pupil of the system in a suitable basis. 
Although other options have also been explored \citep[e.g.,][]{markham99}, the approach
proposed by \cite{paxman92} is very efficient. The second improvement was
the use of phase-diversity techniques \citep{gonsalves79}, which consist of simultaneously
taking pairs of images (or more) with a known static differential aberration.
The third one was a proper treatment of the noise models, which result
in different optimizations. \cite{lofdahl_scharmer94}
and \cite{lofdahl98}, based on the work of \cite{paxman92}, applied it to solar observations, while 
\cite{lofdahl02} extended it to the multiframe case. \cite{vannoort05} later
applied it to the multiobject and multiframe case, developing the
successful multi-object multi-frame blind deconvolution (MOMFBD) code that
is systematically applied in solar filtergraph observations. 
\cite{hirsch11} also considered an online version of multiframe blind deconvolution
that is much more memory efficient.

All previous approaches are conveniently based on the maximization of a proper likelihood
function that is automatically defined by the statistical properties of the noise.
Since maximum likelihood methods can be sensitive to noise, Bayesian approaches are 
also widespread in the blind deconvolution 
community \citep[see, e.g.,][and references therein]{molina01}. This
approach, in which a-priori information about the object or the PSF
is put forward, are arguably more important in single frame deconvolution \cite[e.g.,][]{blanco11}, but they
can be also applied to the multiframe case. For instance, \cite{bucci99} regularize
the phase-diversity problem by imposing additional constraints on the
merit function, while \cite{blanc03} consider the marginal deconvolution
in the same problem. Along this line, \cite{thelen99} solves the blind deconvolution problem
by assuming a multivariate Gaussian prior for the wavefront parameters.

The emergence of deep learning has revolutionized the field of image
processing. In particular, methods have been proposed for the 
deblurring of video sequences \citep{WieHirSchLen17}, for the
rapid estimation of PSFs from images \citep{mockl19} or for the
modeling of simple PSFs \citep{herbel18} for large scale
surveys. It is evident that methods that make use of many frames to produce a single deconvolved frame
make a much better use of the collected photons and should always be 
preferred over lucky imaging techniques. However, their main disadvantage 
resides on the large computational requirements, especially when applied
to long rapid bursts of large images, like in solar observations. Supercomputers 
become necessary to deconvolve the data and deep learning can be  
a remedy to lower the computational requirements. With this idea in mind,
\cite{asensio18} recently developed an extremely fast
multiframe blind deconvolution approach based on supervised deep learning. Although
the model is general, the results were only considered for solar observations. It makes use of
a fully convolutional deep neural network that
was trained supervisedly with images previously corrected with the help of
MOMFBD. Once trained, this method can deconvolve bursts of short-exposure 
1k$\times$1k images in $\sim$ 5 ms with an appropriate Graphical
Processing Unit (GPU). This opens up the possibility, for instance, of doing 
image deconvolution on the fly at the telescope.

Although a step forward in terms of speed, 
the neural approach developed by \cite{asensio18} has two main issues. The
first one is that it is trained with supervision, so one needs to use the
MOMFBD algorithm to build the training set. Though not a major
obstacle, a method that does not need this previous step would be preferable.
The second issue is that the method developed by \cite{asensio18} outputs only
the deconvolved images. No estimation of the wavefront in each individual frame
was produced. Estimating the wavefronts can be helpful to check the performance
of the telescope and instrument, to understand the performance of the
AO or they can be reused when there are several instruments pointing to the
same field of view. For 
these reasons, in this work we present a new deep learning scheme
that can be trained in a fully unsupervised manner, while also
producing an estimation of the wavefront for each observed frame. Given the
lack of supervision, the method can be generally applied to any type of object, either
point-like or extended, once a sufficient number of observed images is available.

\section{Unsupervised training}

\subsection{Image formation}
In this paper, we follow the formalism used by \cite{lofdahl02} and 
\cite{vannoort05}, based on the work of \cite{paxman92}. The deconvolution of a burst 
of short-exposure images\footnote{The exposure time should be small enough
to freeze the atmospheric turbulence in each exposure. In normal seeing conditions, 
an exposure time significantly lower than 10 milliseconds is needed.} is possible once
the linear physics of image formation is imposed. Let us assume that
$o$ is the image of the object under study outside the Earth atmosphere. A burst
of $N$ images taken at times $\{t_1,t_2,\ldots,t_N\}$ through a linear space 
invariant instrument (in our case, telescope+instrument) and corrupted with uncorrelated Gaussian
noise are acquired (see section \ref{sec:loss} for more details). Therefore, 
the image $i_j$ at time $t_j$ that is sensed at the detector 
is given by:
\begin{equation}
   i_j(r) = o(r) * s_j(r) + n_j(r),
   \label{eq:image_formation}
\end{equation}
where $*$ is the convolution operator, $s_j$ is the PSF
of the atmosphere at time $t_j$, $n_j$ is the
uncorrelated Gaussian noise component and $r$ is the spatial coordinate on the
image. Note that the object $o(r)$ is common
to all the $N$ images. Any blind deconvolution method then tries to simultaneously
recover both $o(r)$ and $\mathbf{s}=\{s_1,\ldots,s_N\}$ from the burst of images
$\mathbf{i}=\{i_1,\ldots,i_N\}$. Note that the index $j$ can also be used to refer
to simultaneous images containing known differential aberrations, following the prescriptions of phase-diversity.

The convolution operation in Eq. (\ref{eq:image_formation}) can be translated
into simple multiplications if we transform the equation to the Fourier
space:
\begin{equation}
   I_j(u) = O(u) \cdot S_j(u) + N_j(u),
   \label{eq:image_formation_fourier}
\end{equation}
where the uppercase symbols represent the Fourier transform of the lowercase
symbols and $u$ represents Fourier frequencies. The symbol $S_j(u)$ is known
as the optical transfer function (OTF). The noise is still uncorrelated 
and Gaussian thanks to the linear character of the Fourier transform.

% The space invariant approximation is often violated in normal conditions
% because of the presence of high-altitude turbulence in the atmosphere. This 
% produces different PSFs for different portions of the field-of-view (FOV), with
% sizes defined by the anisoplanatic angle. For this reason, when deconvolving
% extended object, spatially variant PSFs needs to be considered.
% This makes the deconvolution problem much more computationally demanding
% and several approaches have been considered. The simplest one 
% is to solve the deconvolution problems in relatively 
% small overlapping patches which are then 
% stitched together to form the final image.
% This is known as the overlap-add (OPA) approach and is routinely used
% in solar physics with the MOMFBD code with excellent results \cite[e.g.,][]{vannoort05}. 
% More precise and computationally demanding options exist, which are based on the assumption 
% that the PSF has a small domain of support. Among then, one can find
% the widespread method of \cite{nagy_oleary98} and the recent space-variant
% OLA \citep{hirsch10}. We refer to \cite{denis15} for a review of the methods.
% In this work we assume spatially invariant PSFs but our results can
% easily be extended to spatially variant PSFs.

The space invariant approximation is often violated in normal conditions
because of the presence of high-altitude turbulence in the atmosphere. This 
produces different PSFs for different portions of the field-of-view (FOV), with
sizes defined by the anisoplanatic angle. For this reason, when deconvolving
extended object, spatially variant PSFs needs to be considered.
We advise that the overlap-add (OPA) approach is routinely used
in solar physics with the MOMFBD code with excellent results \cite[e.g.,][]{vannoort05}. 
However, more precise approaches like the widespread method of \cite{nagy_oleary98} and 
the recent space-variant OLA \citep{hirsch10} methods can be used \cite[see][for a review]{denis15}.

\subsection{Description of PSFs}
The OTF can be written in terms of the generalized
pupil function:
\begin{equation}
   \label{eq:otf}
   S_j(u) = \mathcal{F}\left[ \lvert \mathcal{F}^{-1}(P_j) \rvert^2 \right].
\end{equation}
In other words, the OTF is the Fourier transform of the PSF which, in turn,
is given by the autocorrelation of the generalized pupil function.
The generalized pupil function can be written as:
\begin{equation}
   \label{eq:generalized_pupil}
   P_j = A_j e^{i \phi_j},
\end{equation}
where $A_j$ describes the amplitude modulation of the pupil (the aperture
of the telescope, including the primary and secondary and any existing
spider) and $\phi_j$ describes the phase at the pupil (also known as wavefront). 
A flat wavefront produces an Airy diffraction PSF. 
The presence of atmospheric turbulence affects this phase by 
introducing a non-flat wavefront which, as a consequence, generates
a complex PSF. Note that this formalism allows us to take into account
a phase-diversity channel by writing down the generalized pupil function as:
\begin{equation}
   P_{j,\mathrm{PD}} = A_j e^{i (\phi_j + \Delta)},
\end{equation}
where $\Delta$ is the added diversity, which is usually a defocus.

Following \cite{paxman92}, we assume that the wavefront can be written (in radians) as a linear 
combination on a suitable basis. The Zernike functions \citep[e.g.,][]{noll76} are among the most useful
functions for reproducing functions in the circle since they 
are orthogonal in the unit circle. 
Despite their nice mathematical properties, Zernike functions are not specially 
suited for efficiently
reproducing wavefronts produced by atmospheric turbulence. The reason is 
that the covariance matrix of the coefficients of the Zernike modes under 
Kolmogorov turbulence (also termed
Noll covariance matrix) is non-diagonal. 
Specifically, the elements of the Noll covariance matrix are given by \citep{roddier90}
\begin{equation}
C_{ij} =\left(\frac{D}{r_{0}}\right)^{\frac{5}{3}}(-1)^{\frac{\left(n_{i}+n_{j}-2 m_{i}\right)}{2}} 
B_{ij} G_{ij}, 
\label{eq:covariance}
\end{equation}
where
\begin{align}
B_{ij} &= \frac{1}{2\pi^2}\sqrt{(n_j+1)(n_i+1)}\Gamma\left(\frac{14}{3}\right) \Gamma\left(\frac{11}{6}\right)^2 \left( \frac{24}{5} \Gamma\left(\frac{6}{5}\right) \right)^{5/6} \\
%\sqrt{\left(n_{i}+1\right)\left(n_{j}+1\right)} \pi^{\frac{8}{3}} \delta_{m_{i}} \delta_{m_{j}}
G_{ij} &=\delta_{m_i,m_j} \frac{ \Gamma\left(\frac{\left(n_{i}+n_{j}-\frac{5}{3}\right)}{2}\right)}{\Gamma\left(\frac{\left(n_{i}-n_{j}+\frac{17}{3}\right)}{2}\right) \Gamma\left(\frac{\left(n_{j}-n_{i}+\frac{17}{3}\right)}{2}\right) \Gamma\left(\frac{\left(n_{i}+n_{j}+\frac{23}{3}\right)}{2}\right)},
\end{align}
where $\Gamma(x)$ is the Gamma function \citep{abramowitz72}, $D$ and $r_0$ are the diameter
of the telescope and Fried radius, respectively. The $\delta_{m_i,m_j}$ Kronecker-like symbol 
is strictly zero when $m_i \neq m_j$ or when $i-j$ is odd (unless $m_i=m_j=0$) and one otherwise.

As a consequence, we use in this paper the so-called Karhunen-Loeve (KL)
modes \citep[e.g.,][]{vannoort05}, which are obtained by numerically diagonalizing the 
covariance matrix given by Eq. (\ref{eq:covariance}). This diagonalization is carried out
using the singular value decomposition, ordering the modes by their singular value. 
Therefore, the wavefront is finally written as:
\begin{equation}
   \label{eq:wavefront}
   \phi_j(x,y) = \sum_{i=1}^M \alpha_{j,i} KL_i(x,y),
\end{equation}
where $M$ is the number of functions used in the linear combination, 
$(x,y)$ refer to coordinates in the pupil plane, $\alpha_{ji}$ are the
$i$-th KL coefficient of the $j$-th wavefront and $KL_i(x,y)$ are obtained as
linear combinations of the Zernike functions.

\subsection{Loss function}
\label{sec:loss}
From a Bayesian point of view, the multiframe blind deconvolution problem requires
the computation of the joint posterior distribution for the 
object $o$ and the $\alphabold=\{\alphabold_1,\alphabold_2,\ldots,\alphabold_N\}$ 
coefficients, conditioned on the observations:
\begin{equation}
   p(o,\alphabold|\mathbf{i}) \propto p(\mathbf{i}|o,\mathbf{s}(\alphabold)) \,
   p(o,\alphabold),
   \label{eq:posterior}
\end{equation}
where the vector $\mathbf{s}(\alphabold)$ refers to the PSFs obtained with the
coefficients $\alphabold$. The posterior distribution is the
product of the likelihood, $p(\mathbf{i}|o,\mathbf{s}(\alphabold))$, and the 
prior, $p(o,\alphabold)$. Note that we make explicit that the likelihood depends on
$\alphabold$ through the PSFs. Sampling the full high-dimensional
posterior is computationally impracticable so point estimates are almost always used. The 
maximum a-posteriori (MAP) solution is therefore given by:
\begin{equation}
   \argmax_{o,\alphabold} p(\mathbf{i}|o,\mathbf{s}(\alphabold)) \,
   p(o,\alphabold).
\end{equation}
Much success has been obtained following this path when dealing with single-image blind 
deconvolution \citep[e.g.,][]{molina01,mistral04,blanco11,babacan12,farrens17,fetick20}. 
We leave the study of this option for a future analysis.

The MAP solution introduces some regularization but it does not
exploit the full potential of the Bayesian approach. Methods based on 
a Type-II maximum likelihood approach require solving the following optimization
problems in marginalized posteriors:
\begin{align}
   \argmax_{\alphabold} \int p(\mathbf{i}|o,\mathbf{s}(\alphabold)) \,
   p(o,\mathbf{s}(\alphabold))\, \mathrm{d}o, \\
   \argmax_{o} \int p(\mathbf{i}|o,\mathbf{s}(\alphabold)) \,
   p(o,\mathbf{s}(\alphabold))\, \mathrm{d}\alphabold.
\end{align}
They arguably lead to better results \citep[see, e.g.,][]{blanco11,fetick20} and
we defer their consideration in the neural unsupervised approach for a future publication.

In this paper we follow the approach of \cite{1994A&AS..107..243L}, \cite{paxman96} and 
\cite{vannoort05} and consider all objects and wavefront coefficients to be 
equally probable a-priori. Taking negative logarithms, 
the maximum likelihood solution that we seek is:
\begin{equation}
   \argmin_{o,\alphabold} L(o,\alphabold),
\end{equation}
where $L(o,\alphabold)=-\log p(\mathbf{i}|o,\mathbf{s}(\alphabold))$ is the negative 
log-likeli\-hood, often termed \emph{loss}
function in the machine learning literature. To simplify the notation, we 
drop the dependency of the log-likelihood on the PSF and show directly
its dependence on the wavefront coefficients. 

\subsubsection{Stationary Gaussian noise model}
Under the assumption of uncorrelated independent and identically distributed
additive Gaussian noise, one can write the loss function as:
\begin{equation}
   L(o,\alphabold) = \sum_r \sum_{j=1}^N \gamma_j \left[ i_j(r) - 
   o(r) * s_j(r;\alphabold) \right]^2,
\end{equation}
where the summation is carried out for all positions $r$ (obviously
now discretized in pixels) of all the
images taken at times $\{t_1,t_2,\ldots,t_N\}$. The
term $\gamma_j$ is an
estimation of the inverse noise variance of the $j$-th image.
Applying both Parseval's and the convolution theorems, one can write
down the log-likelihood with the Fourier components (dropping
unimportant constants):
\begin{equation}
   L(O,\alphabold) = \sum_u \sum_{j=1}^N \gamma_j  \left| H_j(u) I_j(u) - 
   O(u) \cdot S_j(u;\alphabold)\right|^2,
   \label{eq:loss_fourier}
\end{equation}
with the summation now carried out over Fourier frequencies.
Given the historical success of the approach of \cite{paxman96}, we utilize a frequency 
filter $H_j$ to minimize the effect of noise in the observed images. 
The maximum likelihood solution is then formally given by
\begin{equation}
   \argmin_{O,\alphabold} L(O,\alphabold).
\end{equation}
This loss function is non-convex in the set of parameters $\{O,\alphabold\}$, 
and one can apply an alternating optimization method to solve it. This scheme
iteratively considers the two following sub-problems:
\begin{eqnarray}
   \argmin_{O} L(O,\alphabold) \quad \textnormal{with} \,\, \alphabold \,\, \textnormal{constant} \label{eq:subproblem1}\\
   \argmin_{\alphabold} L(O,\alphabold)  \quad \textnormal{with} \,\, O \,\, \textnormal{constant}
\end{eqnarray}
The solution to the linear least squares problem of Eq. (\ref{eq:subproblem1})
is \citep{paxman92}:
\begin{equation}
   \hat{O}(u) = 
        H(u) \frac{\sum_{j} \gamma_j I_{j}(u) S_{j}^{*}(u;\alphabold)}{\sum_{j} \gamma_j \left|S_{j}(u;\alphabold)\right|^{2}}
   \label{eq:image_filter}
\end{equation}
where the caret indicates an estimated quantity. The filter $H(u)$ is used to
ensure that all spatial frequencies for which all OTFs are zero are not used and its specific
form is discussed in Sec. \ref{sec:filter}. 

Another possibility that we explore is to remove the filter from the loss function (\ref{eq:loss_fourier}) and instead
use a Gaussian prior for the object \citep[although it is known that such
priors generate overly smooth images, see][]{tipping02}. The resulting estimation of the object is 
a Wiener filter \citep[e.g.,][]{blanco11}:
\begin{equation}
   \hat{O}(u) = \frac{\sum_{j} \gamma_j I_{j}(u) S_{j}^{*}(u;\alphabold)}{\sum_{j} \gamma_j \left|S_{j}(u;\alphabold)\right|^{2}+\frac{S_n}{S_0(u)}},
   \label{eq:image_wiener}
\end{equation}
where $S_n$ is the power spectral density (PSD) of the noise and $S_0(u)$ is an estimation of the PSD
of the object. The simplest possible Wiener filter assumes $S_n/S_0(u)=K$, with $K$ a constant.

This estimated object can then be 
inserted back into Eq. (\ref{eq:loss_fourier}) resulting in a loss function
that does not depend on the object \citep[see, e.g.,][]{paxman92,vannoort05}. For the
estimation of the object of Eq. (\ref{eq:image_filter}) we have
\begin{equation}
   L(\alphabold)=\sum_u H(u) \left[ \sum_{j} \gamma_j \left|I_{j}(u)\right|^{2}-   
   \frac{\left|\sum_{j} \gamma_j I_{j}^{*}(u) S_{j}(u;\alphabold)\right|^{2}}
{\sum_{j} \gamma_j \left|S_{j}(u;\alphabold)\right|^{2}} \right].
\label{eq:loss_filter}
\end{equation}
For the Wiener filter estimation of the object, we end up with:
\begin{equation}
   L(\alphabold)=\sum_u \left[    
   \frac{\mathcal{S}(\alphabold) \sum_{j} \gamma_j \left|I_{j}(u)\right|^{2}-\left|\sum_{j} \gamma_j I_{j}^{*}(u) S_{j}(u;\alphabold)\right|^{2}}
{\mathcal{S}(\alphabold) + \frac{S_n}{S_0(u)} } \right],
\label{eq:loss_wiener}
\end{equation}
where
\begin{equation}
   \mathcal{S}(\alpha) = \sum_{j} \gamma_j \left|S_{j}(u;\alphabold)\right|^{2}
\end{equation}
In case many objects are observed simultaneously so that they share the same
wavefront, the total loss function is the result of summing the loss function computed for 
each one of the objects.

\begin{figure*}
   \includegraphics[width=\textwidth]{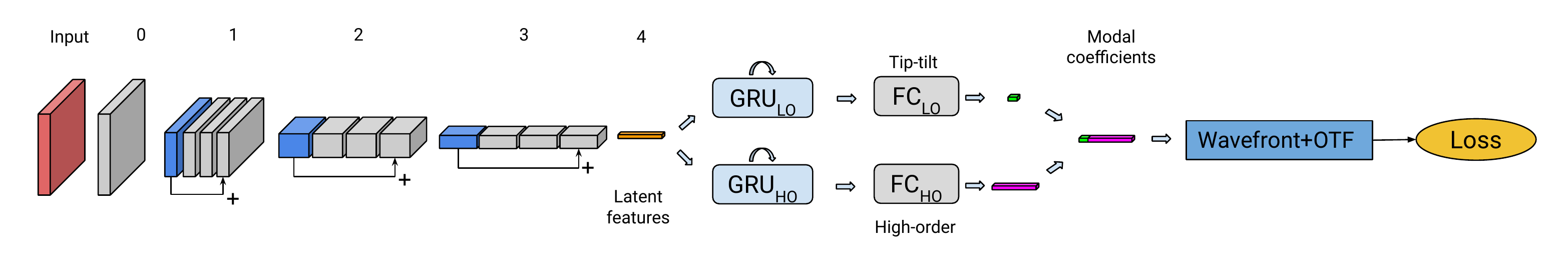}
   \caption{Block diagrams showing the architecture of the network 
   and how it is trained unsupervisedly. The details of each layer are 
   specified in Tab. \ref{tab:modalnet} and in Sec. \ref{sec:network}.
   \label{fig:networks}}
\end{figure*}

\begin{table*}
   \caption{Architecture of encoder-decoder network. The naming convention for 
   the convolutional blocks is $C_{b,l}$, with $b$ referring to the label indicated
   above each block in Fig. \ref{fig:networks} $l$ to the layer inside each block.}
   \label{tab:modalnet}
   \centering
   \begin{tabular}{c c c c c c}
   \hline\hline
   Layer & Type & Kernel size\tablefootmark{a} & Stride & Input tensor shape\tablefootmark{b} & Output tensor shape\tablefootmark{b} \\
   \hline
   $\color{gray} C_{0,1}$ &  BN+ELU+CONV & $3 \times 3 \times C$ & 1 & $W \times W \times C$ & $W \times W \times 32$ \\
   \hline
   $\color{blue} C_{1,1}$ & BN+ELU+CONV & $3 \times 3 \times 32$ & 2 & $W \times W \times 32$ & $W/2 \times W/2 \times 32$ \\
   $\color{gray} C_{1,2}$-\color{gray} $C_{1,4}$ & BN+ELU+CONV & $3 \times 3 \times 32$ & 1 & $W/2 \times W/2 \times 32$ & $W/2 \times W/2 \times 32$ \\
   \hline
   $\color{blue} C_{2,1}$ & BN+ELU+CONV & $3 \times 3 \times 32$ & 2 & $W/2 \times W/2 \times 32$ & $W/4 \times W/4 \times 32$ \\
   $\color{gray} C_{2,2}$-\color{gray} $C_{2,4}$ & BN+ELU+CONV & $3 \times 3 \times 32$ & 1 & $W/4 \times W/4 \times 32$ & $W/4 \times W/4 \times 32$ \\
   \hline
   $\color{blue} C_{3,1}$ & BN+ELU+CONV & $3 \times 3 \times 32$ & 2 & $W/4 \times W/4 \times 32$ & $W/8 \times W/8 \times 32$ \\  
   $\color{gray} C_{3,2}$-\color{gray} $C_{3,4}$ & BN+ELU+CONV & $3 \times 3 \times 32$ & 1 & $W/8 \times W/8 \times 32$ & $W/8 \times W/8 \times 32$ \\
   \hline
   $\color{orange} C_{4}$ & CONV & $W/8 \times W/8 \times 512$ & 1 & $W/8 \times W/8 \times 32$ & $1 \times 1 \times 512$ \\  
   \end{tabular}
   \tablefoot{
   \tablefoottext{a}{Kernel dimensions: $W \times W \times C$: W: kernel horizontal size, C: kernel depth.}
   \tablefoottext{b}{Image dimensions: $W \times W \times C$: W: image horizontal size, C: image number of channels.}
   }
   \end{table*}

Eqs. (\ref{eq:loss_filter}) and (\ref{eq:loss_wiener}) define loss functions that can be optimized with respect
to $\alphabold$ to find the instantaneous wavefront. From these coefficients, the PSFs 
affecting each one of the $N$ frames of the burst can be computed.
Once the wavefronts are computed, the deconvolved image can be easily
obtained using Eqs. (\ref{eq:image_filter}) or (\ref{eq:image_wiener}), although
more elaborate non-blind deconvolution solutions can also be utilized. We use Eq. (\ref{eq:image_filter})
for all the results shown in this paper, so that:
\begin{equation}
   O = P_{+} \left[ \mathcal{F}^{-1}(\hat{O}(u)) \right],
\end{equation}
where we also enforce non-negativity by using the $P_+$ operator that 
sets all negative pixels to zero.

% Since Eq. (\ref{eq:image}) represents an infinite amount of
% maximum likelihood images once the Hermitian property is fulfilled
% in $\chi_0$, obtaining a suitable image requires some work. One option
% is to set a non-flat prior $p(o)$. This is the approach followed
% by many works dealing with single-image blind 
% deconvolution \citep[e.g.,][]{molina01,mistral04,blanco11,fetick20}. 
% We leave the study of this option for future analysis since, a priori, one needs to make sure  
% that the prior allows an analytical solution to Eq. (\ref{eq:subproblem1}) if one
% wants to follow the strategy described in this paper.
% A case that would fulfill this condition is a Gaussian prior, but they 
% are known to generate overly smooth images \citep{tipping02}. However, we also believe
% that a more complex neural architecture can deal with the case in which the loss function
% is explicitly dependent on the image $o$. For instance, a new branch on the
% neural architecture can be built to give an estimation of this deconvolved image. This
% architecture could potentially deal with any desired image prior.

\subsubsection{Other noise models}
The dominant noise in astronomical imaging is photon noise, which
follows a Poisson distribution. The main difficulty in using the Poisson
log-likelihood in our scheme is that a closed form solution to 
Eq. (\ref{eq:subproblem1}) has not been found yet \citep{paxman92}. Fortunately,
the Gaussian distribution is a good approximation to the Poissonian once the
number of photons is larger than $\sim$20. For this to happen, one has to 
make the distribution non-stationary with a variance equal to the number
of photons, which avoids an easy transformation to the Fourier
space.

Anyway, the observations that we use for training are not in the low-photon limit
and the sources are not very bright with respect to the background. As a consequence,
and since our objective is to show how a neural network can be trained unsupervisedly
for performing multiframe blind deconvolution, we stick to the Gaussian case with
constant $\gamma_j$ per frame.

\subsubsection{Filter}
\label{sec:filter}
The visual appearance of the deconvolved image heavily depends on the specific
details of the non-blind deconvolution technique that we use once the wavefront
coefficients are obtained. The results shown in this paper have been
computed using Eq. (\ref{eq:image_filter}), with the filter 
proposed by \cite{lofdahl94} that has shown very good practical results 
in solar images \citep{lofdahl02,vannoort05}. The filter has the following
form:
\begin{equation}
   H = 1 - \frac{\sum_{j} \left|S_{j}\right|^{2}}{\left| \sum_{j} I_{j} S_{j}^{*} \right|^2},
\end{equation}
where we set to zero all values below 0.2 and above 1. Finally, we remove all isolated
peaks in the filter that cannot be directly connected to the peak at zero frequency with $H>0.2$.

\subsection{Data preprocessing}
Apart from the standard ill-definition of the multiframe blind deconvolution problem, that 
can be alleviated with prior information, this method is always subject to some 
fundamental ambiguities that are harder to deal with \citep{paxman19}. One of the most 
critical ones in our approach is the fact that the global
tip-tilt cannot be obtained. If the object is shifted by a certain amount, one can always
compensate for it with a tip-tilt contribution in the wavefront so that the 
image of the object remains stationary. Consequently, any learning method that
we use will get confused on the specific amount of tip-tilt to infer from the
images. We force the results to have small tip-tilt coefficients by pre-aligning all images 
of the burst so that the object of interest is, on average, centered on the field of view.
We do this by computing the sum of all the images in the burst, computing the peak emission
and shifting this peak to lie at the center of the image with pixel precision.

\subsection{Neural architecture}
\label{sec:network}
Our neural approach is based on the construction of a neural network that can
directly predict the vector $\alphabold$ from the images of the burst.
This architecture is broadly made of the following components. The first is a convolutional neural network 
shared for all frames of the burst that extracts features from individual images of 
size $W\times W$. Two recurrent neural network (RNN) follow and they take into
account the time evolution of the tip-tilt and high-order coefficients. The RNN in charge
of the tip-tilt is not applied to the first frame, which is assumed to have zero tip-tilt. The
main purpose of this RNN is to give the relative tip-tilt with respect to the first
frame. The high-order RNN is applied to all frames. Two fully connected neural network (FC$_\mathrm{LO}$ and FC$_\mathrm{HO}$)
is shared by all frames and produces two heads for the prediction of tip-tilt and high-order aberrations. 
The two predictions are concatenated and, finally, we find a layer that computes the OTFs from the 
wavefront coefficients, which are then plugged into the loss function of Eq. (\ref{eq:loss_filter}) for training
end-to-end the architecture. Our approach is graphically depicted 
in Fig. \ref{fig:networks} and in the following we describe each component in detail.

% \begin{figure}[!b]
   % \includegraphics[width=\columnwidth]{figs/NOT_pupil.pdf}
   % \caption{Example of the turbulent phase, measured in radians, at the NOT pupil produced by a 
   % realization of a Kolmogorov synthetic atmosphere with a Fried radius of $r_0=10$ cm.}
   % \label{fig:pupil}
   % \end{figure}
   
\subsubsection{Convolutional neural network}
The aim of the first element of the architecture is to summarize the 
images and extract all relevant information in a vector, that can
be used later for the prediction of the wavefront coefficients. This
component is shared among all frames, so it can be applied in parallel for 
all the inputs images. This neural network is a fully convolutional encoder, whose properties
are summarized in Tab. \ref{tab:modalnet}. The first step is a convolutional
layer with a $3 \times 3$ kernel that generates 32 channels from the 
input tensor. Both in the application to point-like objects and to
extended objects, we consider an input tensor with several channels. They
will be explicited in Sec. \ref{sec:point} and \ref{sec:extended}.
Then, a series of standard convolutional blocks made of the consecutive
application of batch normalization \citep{ioffe_batchnormalization15}, an exponential linear
unit activation function \citep[ELU;][]{clevert_elu15} and a convolutional layer with
the kernel size specified in Tab. \ref{tab:modalnet} are applied, generating intermediate
feature tensors. In order to accelerate convergence,
skip connections are added between the initial layer of a block and the last one. A final layer,
indicated in orange, uses a kernel of size $W/8\times W/8$ to produce a vector of size
512 as output. Note that the input images need to have a size multiple of $2^3$ to produce an integer
number of latent features.

\subsubsection{Recurrent neural network}
The aim of the RNNs are to estimate the tip-tilt and high-order coefficients of the wavefront
for all images of the burst while keeping some memory from the rest of frames. The RNNs
can exploit any existing temporal correlation, consequence of the rapid cadence of the observations.
Additionally, it can potentially understand that all frames share the same object, which can be
helpful for a better estimation of the aberrations.

Although it is not possible 
to detect an absolute tip-tilt for a single image, we can use the
first frame as a reference and estimate the tip-tilt of all remaining frames relative
to the first one. Both RNNs are Gated Recurrent Units \citep[GRU;][]{gru14}, which
are able to deal with relatively long sequences. We have also experimented
with Long-Term Short Memory units \citep{lstm97} with good results, although they are more computationally
demanding. GRUs contain an internal state (\emph{cell}) that remembers values over 
long sequences, and gates (\emph{reset} and \emph{update})
that are used to control the flow of information into and out of the cell.
We choose the cell to be a vector of length 512, of the same length as the input. We also choose
the GRU to be bidirectional, so it attends to the inputs in the two possible directions, from frame 1 to frame
$N$ and vice\-versa.

\subsubsection{Fully connected neural networks}
Two fully connected neural networks produce the final estimation of the tip-tilt
and high-order aberrations. The layers are defined by the consecutive application of
two linear transformations of sizes $512 \times 512$ and $512 \times 512$,
each one followed by an ELU activation. The two heads are obtained by
predicting the tip-tilt with a final linear transformation of size $512 \times 2$ and 
the remaining high-order aberrations with a linear transformation
of size $512 \times (M-2)$. The predicted wavefront coefficients are computed by
applying a final activation function $a \tanh(b x)$. After some trial-and-error, we found
that $a=2$ and $b=1/10$ gave good results. Since the output is limited to the interval $[-a,a]$,
one can increase $a$ if very large aberrations are expected.

\subsubsection{Computation of OTFs}
Once the wavefront coefficients are known for all images in the burst, 
one can use Eq. (\ref{eq:wavefront}) to compute the phase on the pupil. Then,
the generalized pupil function is obtained from Eq. (\ref{eq:generalized_pupil})
and the OTF from Eq. (\ref{eq:otf}). This, together with the Fourier transforms
of the input images, are all the ingredients needed for the computation of
the loss function using Eq. (\ref{eq:loss_filter}).

\subsection{Training}
\label{sec:modalnet}
The training is done by modifying the parameters of the neural networks 
so that the loss function of Eq. (\ref{eq:loss_filter}) is minimized 
for a suitable training set. The several components of our architecture
have a total number of $\sim$7.7 M free parameters. The training is
carried out using backpropagation, i.e., computing the derivative of
the loss function with respect to the free parameters and using this 
gradient to modify them. The recurrent neural network needs to be trained
using backpropagation in time. To this end, it is unrolled for 25
steps and considered it as a normal fully connected neural
network.

\section{Results: point-like objects}
\label{sec:point}
As a first step, we consider point-like objects. They are not really the 
main subject of deconvolution methods since their properties (e.g., astrometry) can be measured
with other methods \citep[e.g.,][]{WEIGELT197755}. However, they can be
useful to check whether the estimated wavefronts are representative of the 
instantaneous PSFs.

\subsection{Baseline}
In order to check the ability of out neural approach to correctly estimate the
wavefront coefficients, we compare them with a standard multiframe blind deconvolution method. This
baseline is obtained by minimizing the loss function of Eq. (\ref{eq:loss_filter}) using the KL coefficients
of the wavefront in each frame as unknowns. We use PyTorch to optimize this loss function using the
Adam optimizer with a learning rate of 0.1. This learning rate was selected by trial and error.
The average computing time per iteration for the deconvolution
of 100 frames is $\sim$0.8 s. The typical number of iterations for convergence is around 70, so
the deconvolution can be achieved in around one minute. Obtaining
reliable results for sources with reduced signal-to-noise ratios (SNR) per frame turned out to be challenging and, in some
occasions, impossible.

\subsection{Training set}
% \begin{figure}[!b]
%    \centering
%    \includegraphics[width=0.9\columnwidth]{figs/SIGORI_classic_deconvolved.pdf}   
%    \includegraphics[width=0.9\columnwidth]{figs/GJ661_classic_deconvolved.pdf}
%    \includegraphics[width=0.9\columnwidth]{figs/GJ569_classic_deconvolved.pdf}      
%    \caption{Deconvolved images with the classical approach using 100 frames
%    for three different sources.}
%    \label{fig:classic}
%    \end{figure}

For the examples shown in this section we choose observations carried out with 
the FastCam instrument 
mounted on the Nordic Optical Telescope (NOT) on the Observatorio del Roque
de los Muchachos (La Palma, Spain). 
FastCam is a lucky imaging instrument jointly developed by the 
Spanish Instituto de Astrof\'{\i}sica de Canarias and the Universidad 
Polit\'ecnica de Cartagena. The instrument uses an Andor iXon DU-897 
back-illuminated EMCCD containing a 512x512 pixel frame. 
The observations were carried out with
a standard I Johnson-Bessel filter at an effective wavelength of 824 nm 
with a with of 175 nm. The pixel size was 0.0303". The telescope diameter
is 2.56 m, with a central obscuration of 0.51 m, giving a diffraction limit
of 0.0786". The observations were obtained on four 
consecutive nights on 2007 October 3-6, 
and they include the following objects: GJ1002, GJ144, GJ205, GJ661, RHY1, RHY44,
for a total of several hundred thousand images of 128$\times$128 pixels during the four-days run. 
Some of them are single stars in the FOV and others contain a pair of
stars. The training set consists of 40 bursts of 1000 images each with an exposure time of
30 ms, enough to efficiently freeze the atmospheric turbulence. The images
are taken at different times, and they cover reasonably variable seeing conditions. 
A validation set of 9 bursts, not used for training, is put apart to check for overfitting.
Given the unsupervised character of our approach, the neural network can be easily refined by adding more observations
which can cover different seeing conditions.

\begin{figure*}
   \includegraphics[width=0.5\textwidth]{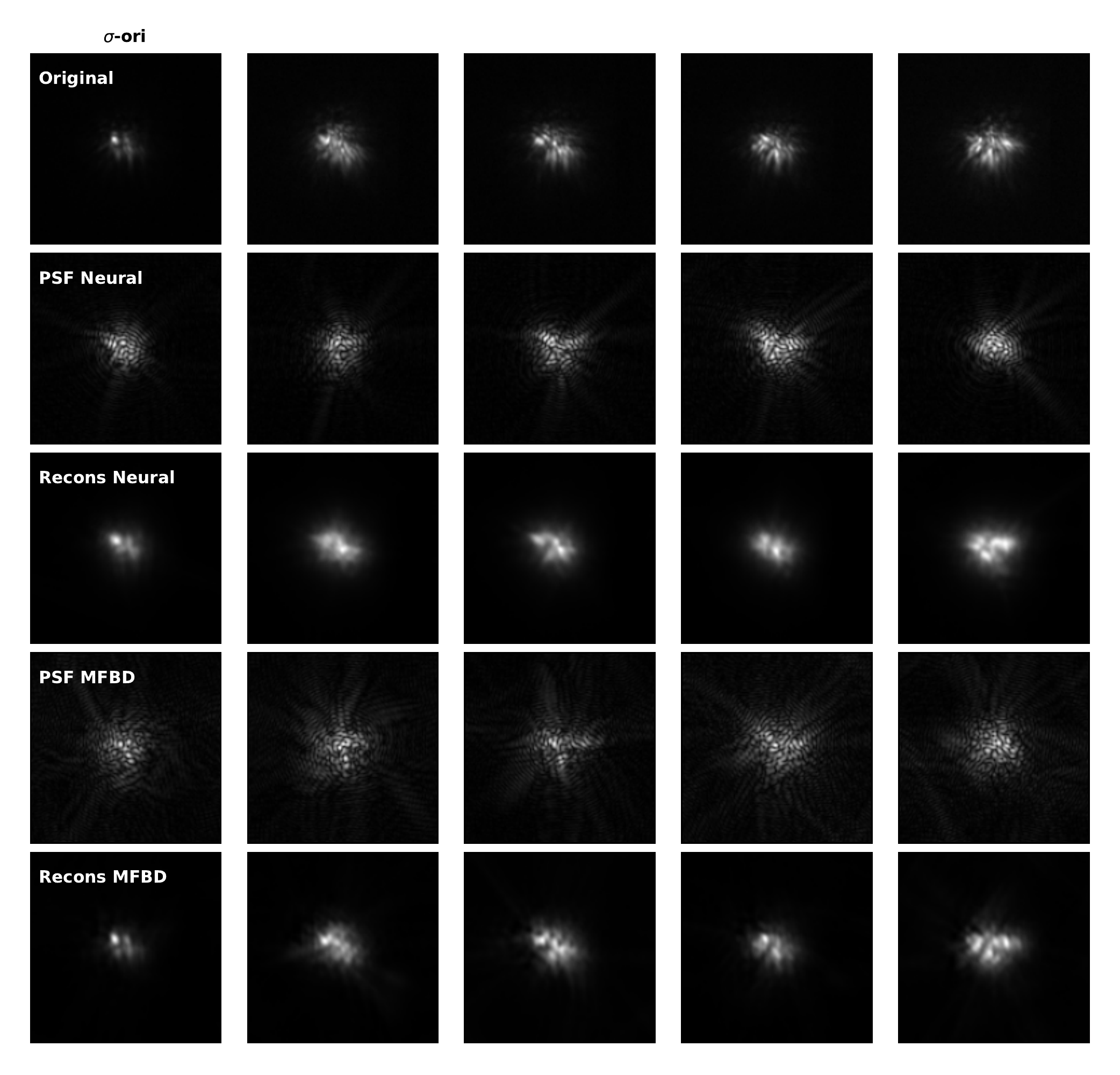}
   \includegraphics[width=0.5\textwidth]{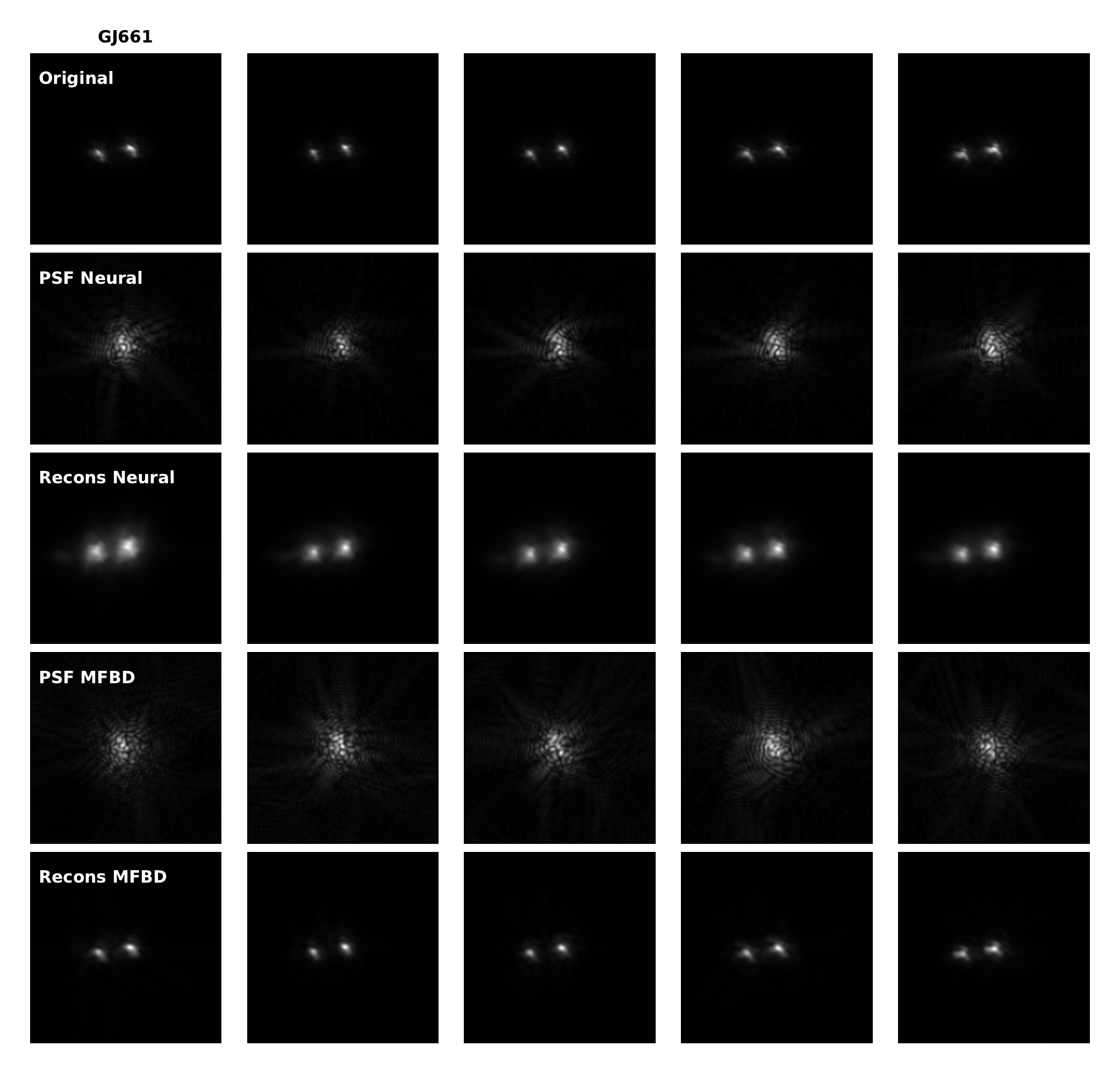}\\
   \includegraphics[width=0.5\textwidth]{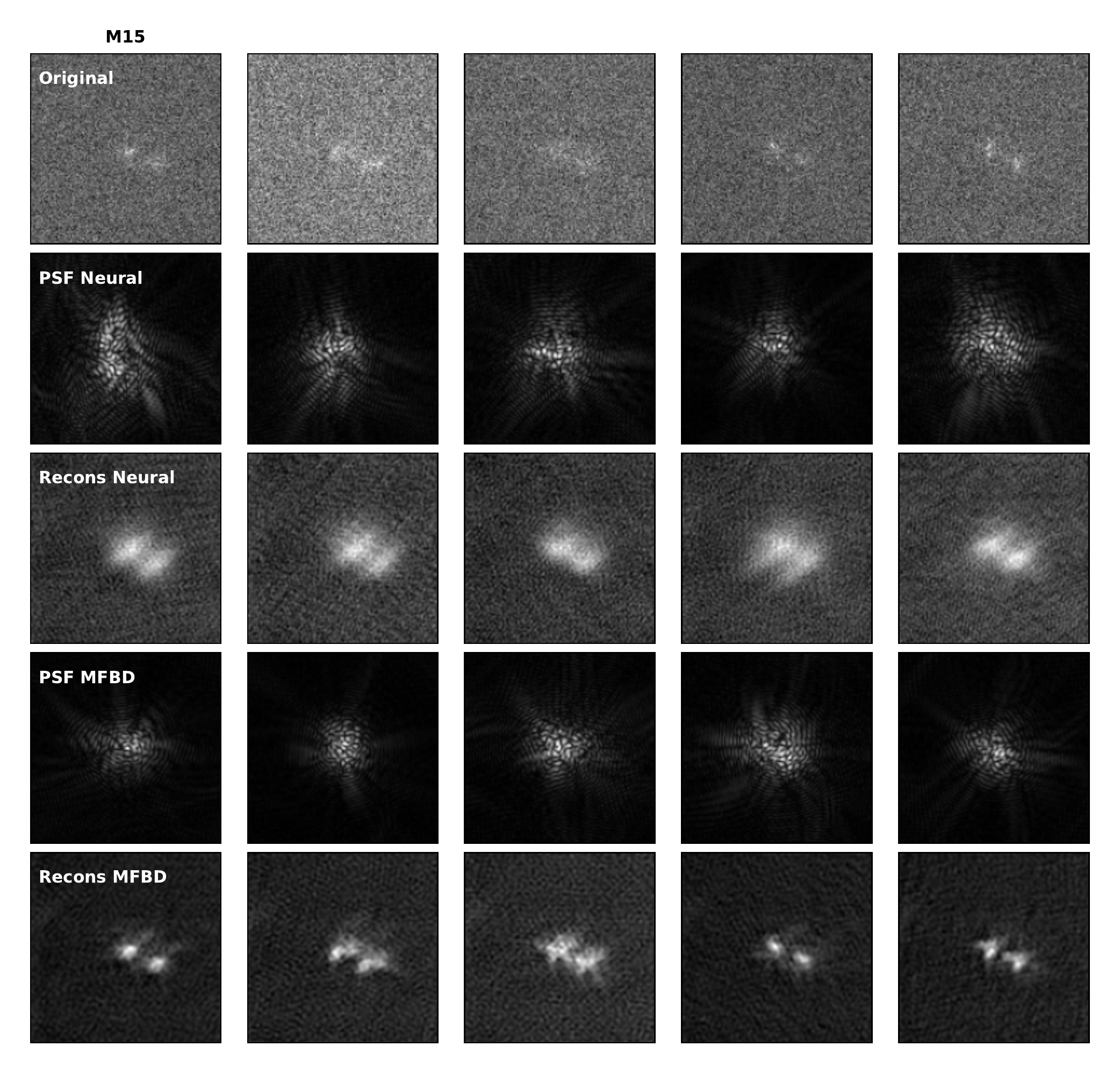}
   \includegraphics[width=0.5\textwidth]{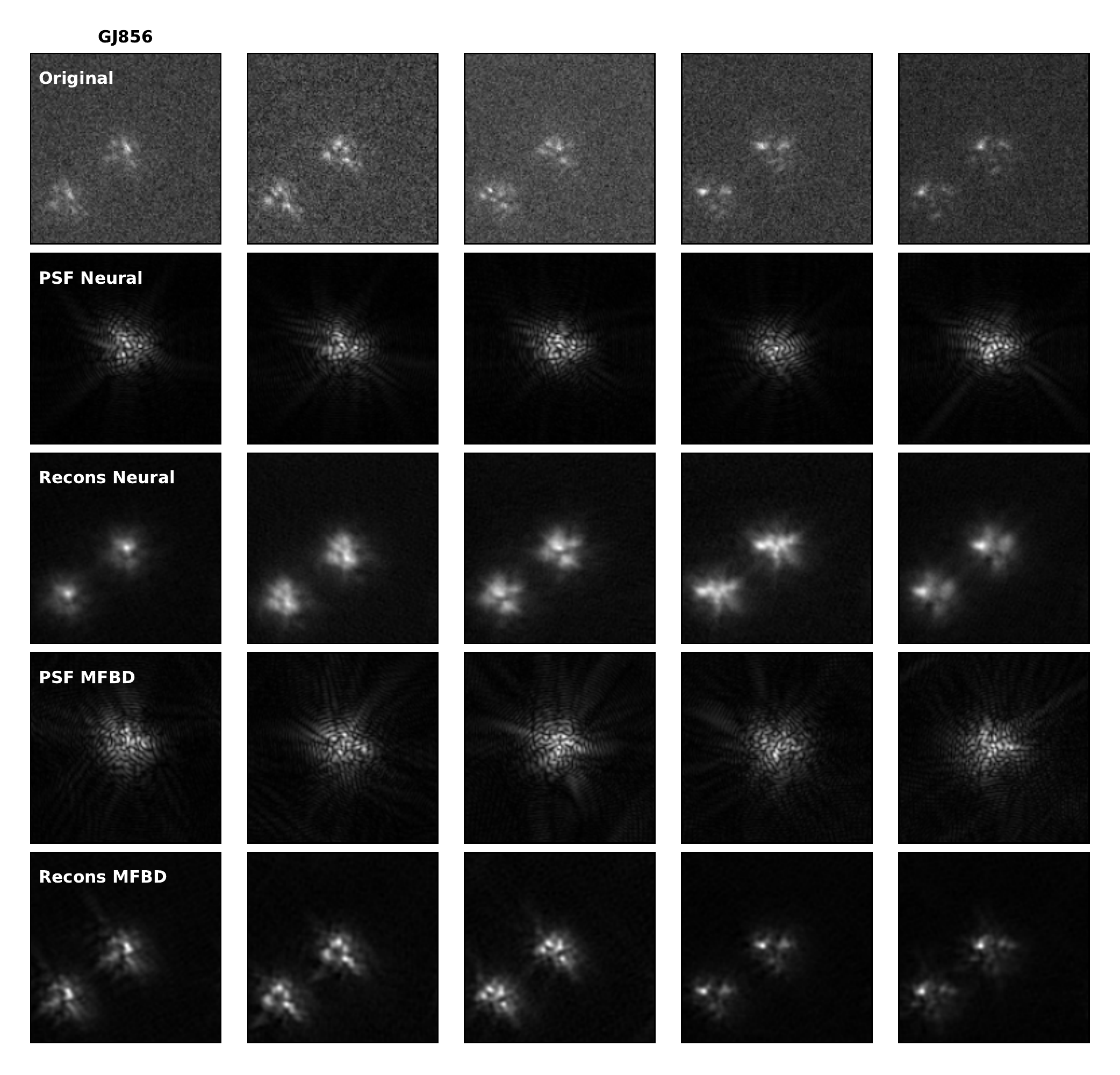}
   \caption{Original frames of the burst (first row), estimated PSF with the neural
   approach (second row) and the baseline (fourth row), together with the reconstructed
   individual images (third and fifth rows). Several sources with different seeing conditions
   and characteristics are displayed.}
   \label{fig:examples_stars}
   \end{figure*}

The training is done by randomly extracting 1000 short bursts of 25
frames (this is the number of unrolled steps of the GRU recurrent component
of our architecture) from each one of the 40 available observations, for a total of 40000 training
examples. To facilitate the training, the images are normalized by computing the maximum
and minimum in the burst and mapping these values to the $[0,1]$
interval. Additionally, we use two channels as input, one containing the normalized image itself, 
that can be useful for inferring properties of the center of the PSF, and the other one containing 
its square root, which gives a better contrast to the tails of the PSF

Once the wavefront coefficients are computed, this normalization
is not needed and the deconvolved image can be reconstructed using
the original images.

   % \begin{figure*}
   %    \centering
   %    \includegraphics[width=0.9\textwidth]{figs/M15_psfs.pdf}
   %    \includegraphics[width=0.9\textwidth]{figs/GJ856_psfs.pdf}
   %    \caption{Same as Fig. \ref{fig:examples1}, but for M15 in the upper panel and
   %    GJ856 in the lower panel.}
   %    \label{fig:examples2}
   %    \end{figure*}

   \begin{figure*}
      \centering
      \includegraphics[width=0.85\textwidth]{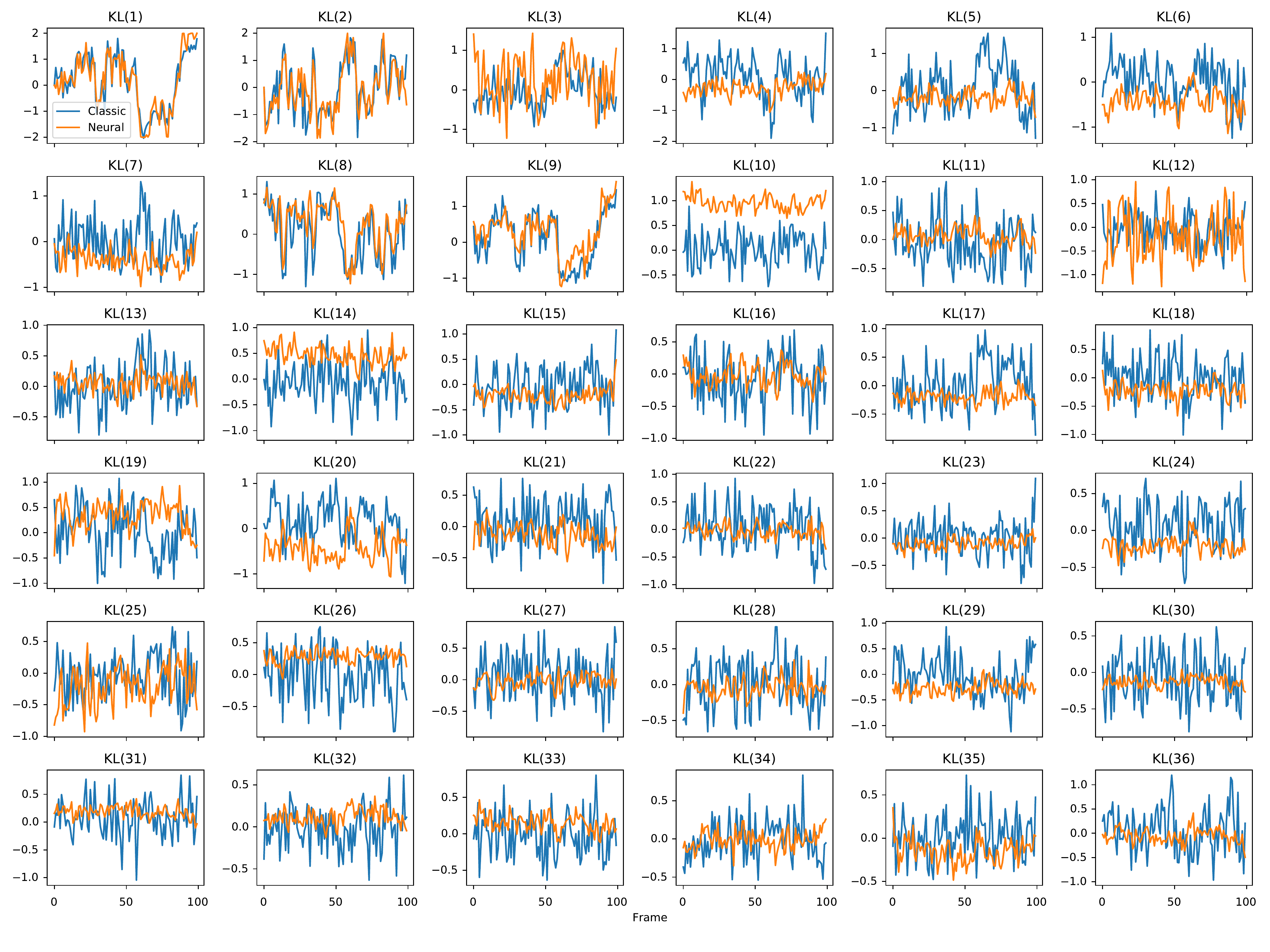}
      \caption{Time evolution of the first 36 coefficients of the phase in the KL
      basis in radians for the $\sigma$-Ori observation. The parameters found by the
      classical MFBD solution are shown in blue, while those inferred by the neural 
      network are shown in orange.}
      \label{fig:kl}
      \end{figure*}
   
The following augmenting strategy is applied to help improve the generalization 
capabilities of the neural network. Each 
burst is randomly rotated 0, 90, 180 or 270 degrees and 
flipped horizontally or vertically with equal probability. 
The neural networks are implemented in PyTorch 1.6 \citep{pytorch19}. We use the
Adam optimizer \citep{adam14}, with a learning
rate of 3$\times$10$^{-4}$ and a batch size of 16, during 25 epochs.
We found that the chosen learning rate produces 
suitable results and it was kept fixed for all experiments.  
As a consequence of the serial character of the recurrent network, each epoch takes
roughly $\sim$17 min, so the total training time is roughly 7 hours on a single NVIDIA RTX 2080 Ti GPU.
A large fraction of the computing time in the forward pass occurs on the computation of the OTFs and the
computation of the loss function.

When in evaluation mode, the output of the network are the wavefront coefficients. Unless one
is interested in computing the deconvolved image, the calculation of the OTF can be safely ignored.
Perhaps the largest difference with respect to the direct optimization of the loss function
is that the computing time for a single pass for 100 frames is 20 ms,
almost 3000 times faster. This time includes the input/output time to/from the GPU and
contains some overheads that can be easily avoided. Additionally, thanks to the inherent 
parallelization in GPUs, the time per deconvolution can be reduced if several stars are deconvolved 
concurrently. The only limitation is the amount of memory on the GPU.

\subsection{PSFs}
Figure \ref{fig:examples_stars} shows examples of the inferred PSFs for different stars
from the test set. We show the first five frames of the burst in the upper row. We can immediately verify
that the seeing conditions and SNR are different in 
all the examples we consider. For instance, the spread of the images in $\sigma$-Ori is much 
larger than that of GJ661, indicating a larger turbulence.

The second row of the panels displays the instantaneous PSF (in fact we display the square root
to increase the dynamic range of the plot) estimated by the neural 
approach, while the fourth row shows the results of the baseline. 
The results clearly show that, in general, we are capturing the shape of the PSF correctly, including 
a large fraction of the wings. After extensive experiments, we have not found any clear sign of 
PSF degeneracy, fundamentally a consequence of assuming a pupil based PSF.

Finally, as a consistency check, we re-convolve the image obtained from Eq. (\ref{eq:image_filter})
with the estimated PSF in both the neural and baseline cases. Our aim with this experiment 
is to give a visual approximate cross-check of the quality of the 
inferred PSF. Obviously, the resulting images should be similar to each observed frame, apart from the obvious noise
reduction consequence of the cleaner deconvolved image. However, one should be
cautious because the result strongly depends on the quality of the deconvolved
image, which crucially depends on the filter in the case of Eq. (\ref{eq:image_filter})
or the estimated PSD of the object in the case of Eq. (\ref{eq:image_wiener}).
Anyway, one can see minute details of the image that are reproduced with great fidelity in the re-convolved
image. Perhaps one can argue that, in cases of very bad seeing with complex PSFs like
the case of $\sigma$-Ori, the re-convolved object is slightly more diffuse than the original one.
However, the neural approach captures enough details of the PSF so that the ensuing
deconvolved image can be made of very high quality.

\begin{figure*}
   \includegraphics[width=0.5\textwidth]{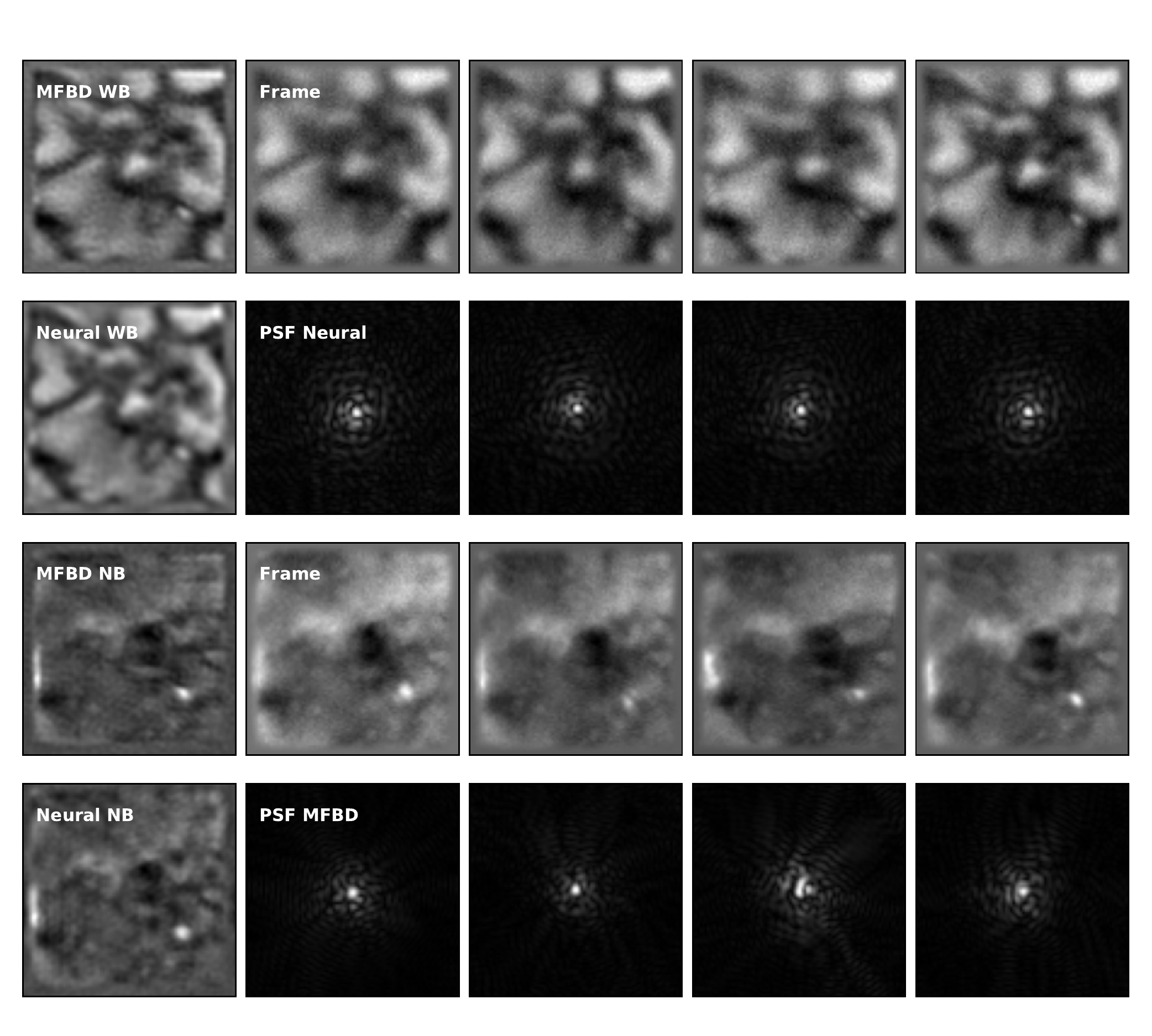}
   \includegraphics[width=0.5\textwidth]{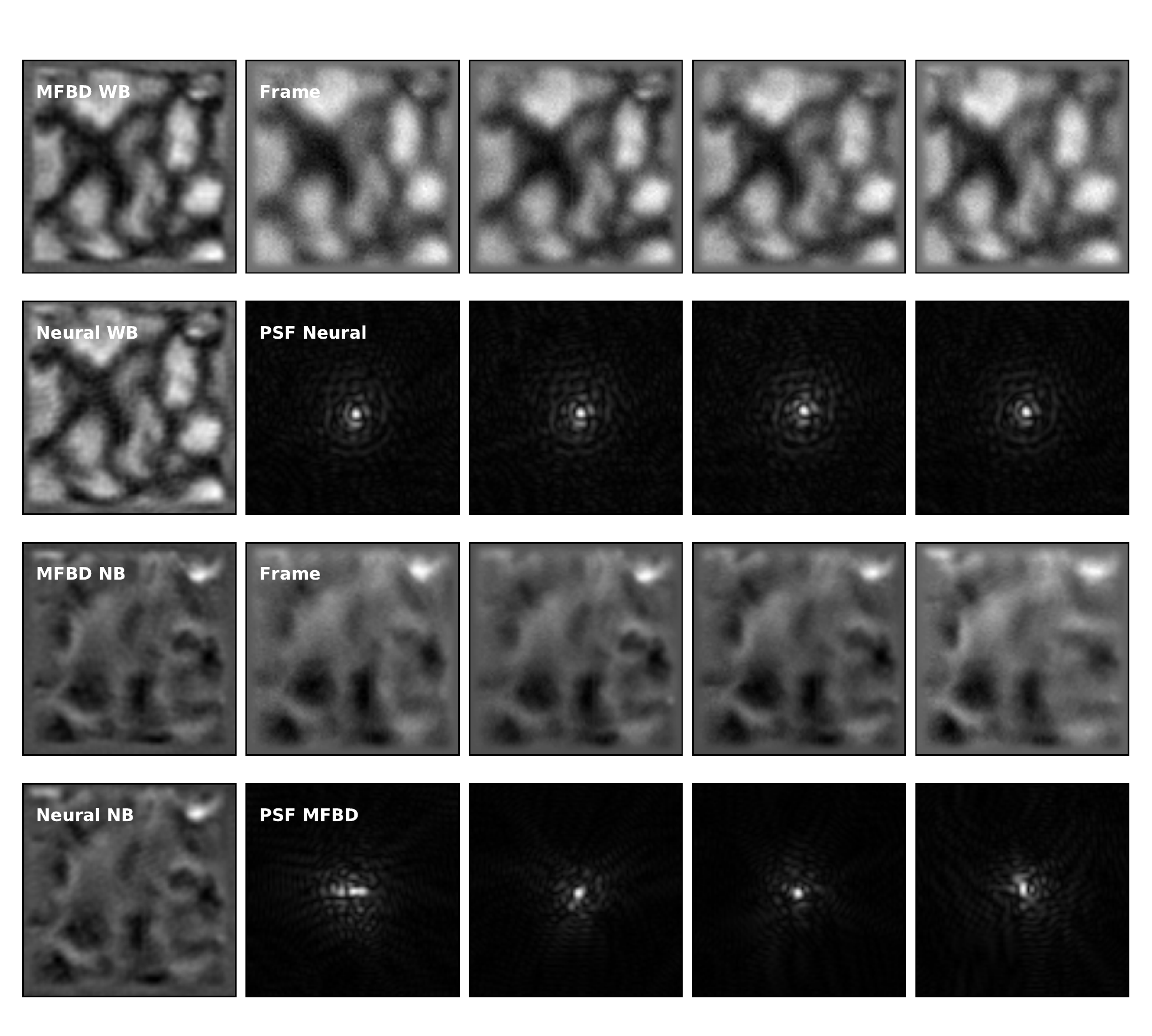}\\
   \includegraphics[width=0.5\textwidth]{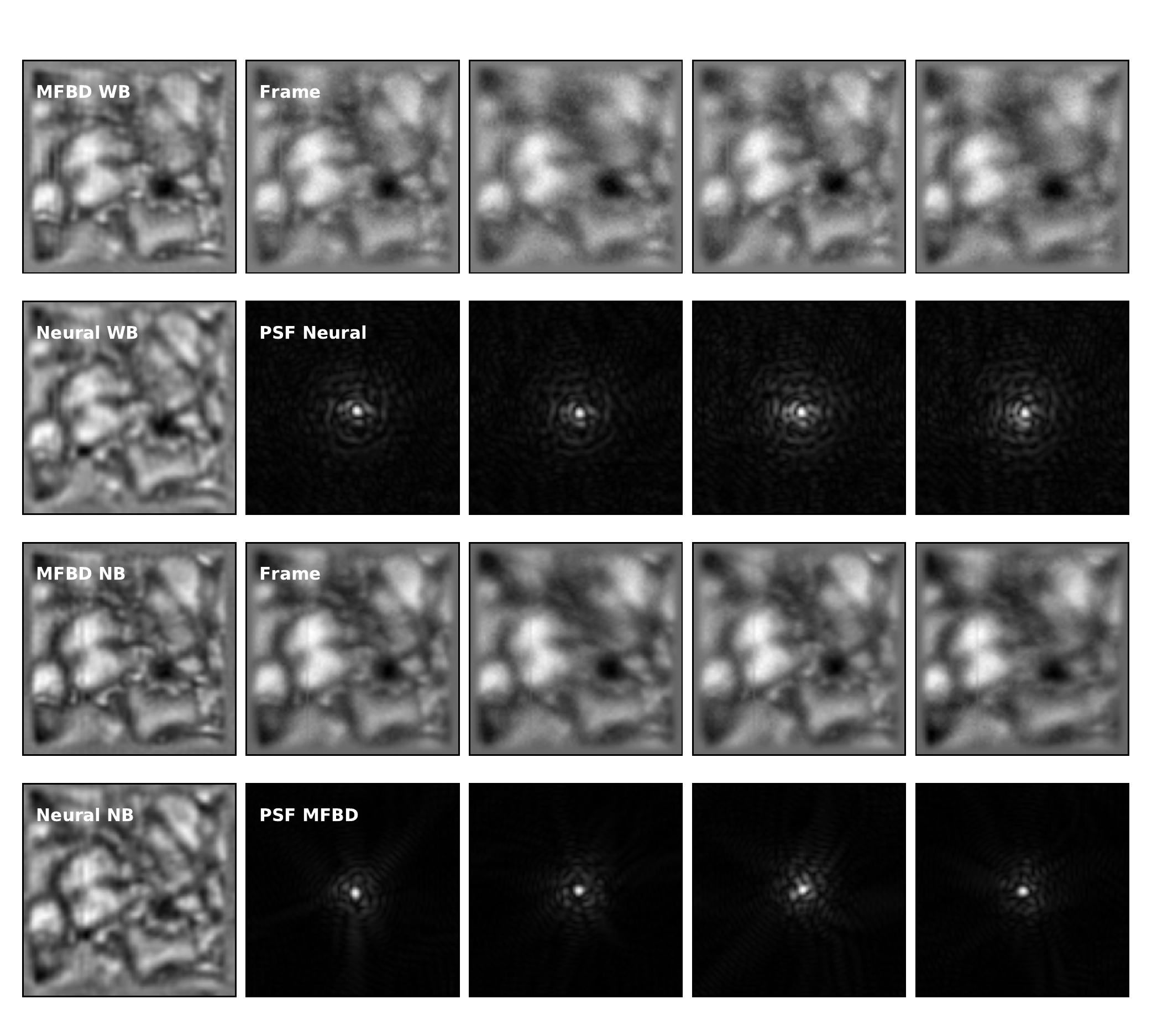}
   \includegraphics[width=0.5\textwidth]{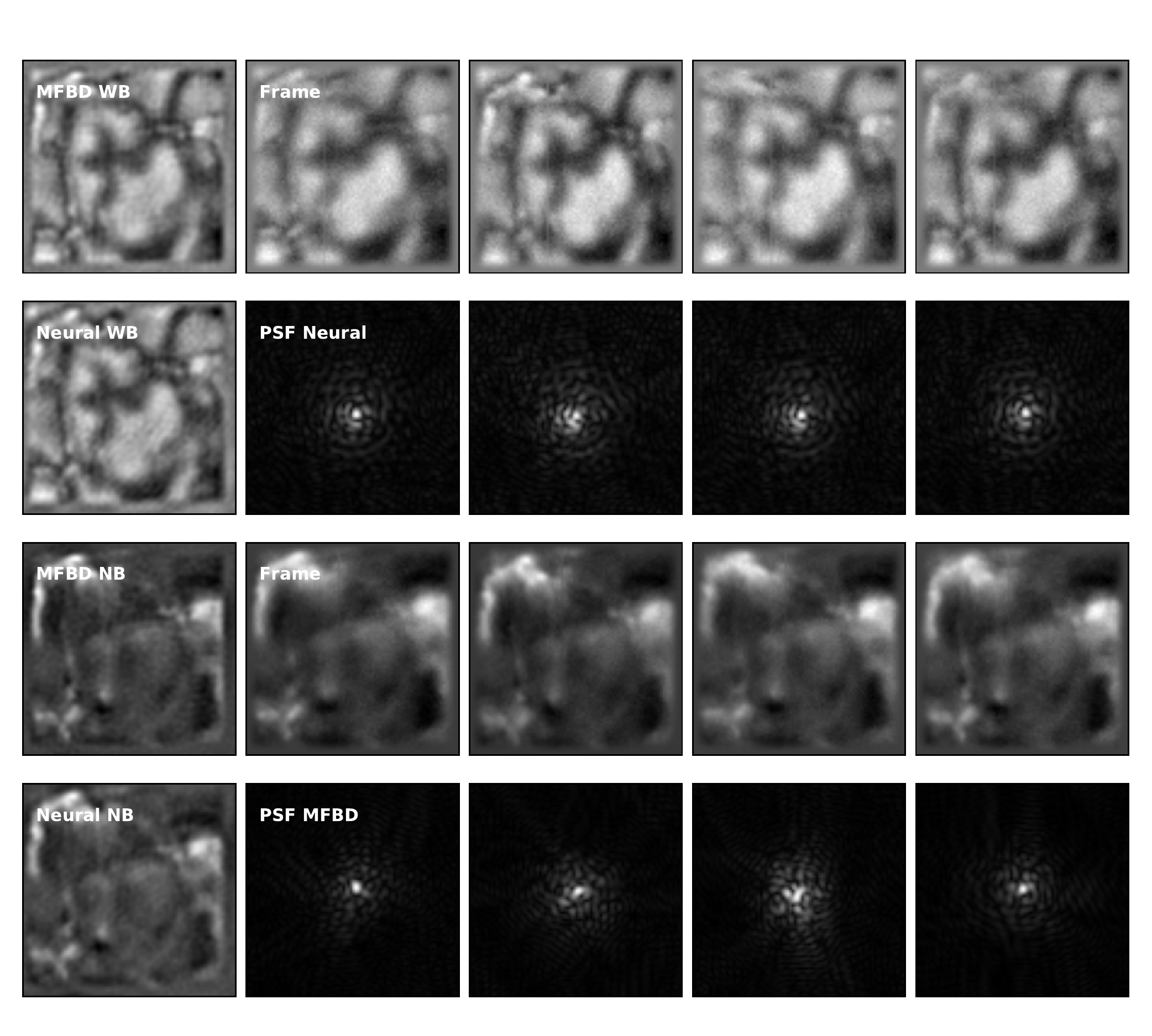}\\
   \caption{Four examples of extended objects deconvolution, each one showing results for the
   WB and one of the NB channels. The upper row displays the baseline deconvolution, together with four 
   raw frames. The second and fourth rows display the estimated PSFs by the neural network
   and the baseline, respectively. Since NB and WB channels are simultaneous, the PSFs are shared
   for the two objects.}
   \label{fig:examples_sun}
   \end{figure*}

\subsection{Karhunen-Loeve coefficients}
A different way of comparing the baseline and our neural approach is by analyzing the
inferred wavefront coefficients. Figure \ref{fig:kl} shows the first 36 KL modes 
for $\sigma$-Ori, in blue the baseline and in orange those obtained with the neural
network. These results are relevant because the results of the baseline have been used by
\citep{2017A&A...608A..76V} to correct strictly simultaneous spectropolarimetric
data with excellent results. Consequently, our approach can produce fundamentally
the same instantaneous PSFs with a much reduced computational burden. Apart from that,
it turns out relatively trivial to obtain instantaneous properties of the PSF (e.g., Strehl ratio)
from these coefficients.

The tip-tilt coefficients (first and second coefficients) are very well
estimated. The same happens with $KL_8$ and $KL_9$. There are some discrepancies in
some of them, especially in $KL_{10}$. In general, we find that the high-order 
coefficients are correctly obtained on average, but their amplitude is short in comparison
with those of the baseline. Anyway, there are potential quasi-ambiguities in the problem
for which different combinations of KL modes can produce very similar PSFs that cannot
be distinguished during the deconvolution. The recurrent
structure in our neural architecture is able to exploit the time correlation 
that is present in the wavefront coefficients. Note that, although each frame 
has 30 ms exposure time, the overhead due to readout is $\sim$56\%. Therefore, the total elapsed
time for 100 frames is roughly 4.7 s.

\section{Results: extended objects}
\label{sec:extended}
The previous results show that it is possible to unsupervisedly train our
system to estimate wavefronts from burst of images of stellar objects. It is 
true that the PSF is directly accessible from the image when dealing with 
point-like objects, even though it might be repeated several times in the FOV because
of the presence of several objects in the FOV. The extended case that we face in this 
section is much more challenging because
the neural network has to be able to estimate decent wavefront coefficients from
images that fill the FOV and have arbitrary brightness variations.

\subsection{Training and validation sets}
We employ the same datasets that were used in \cite{asensio18}. They were 
observed with the CRisp Imaging SpectroPolarimeter (CRISP) instrument at 
the Swedish 1-m Solar Telescope (SST) on the Observatorio del Roque de los Muchachos (Spain). 
The data used for training are spectral scans on the \ion{Fe}{i} doublet 
on 6301-6302 \AA, containing 15 wavelength points with a pixel size of $0.059"$. The 
observations include the 4 polarization
modulation states that are used to measure the full-Stokes vector. 
The polarimetric modulation is carried out at each wavelength sequentially, producing 
two narrow band (NB) orthogonal images (CRISP uses dual beam polarimetry to minimize
seeing-induced cross-talk) for a set of 7 acquisitions of 17.35 ms each. Additionally,
a wide band (WB) image is strictly synchronized with the NB images. The three images are
used as input in the neural network. The images of the training set are corrected 
following the standard procedure \citep{jaime15}, that includes: dark current subtraction,
flat-field correction, and subpixel image alignment between the two NB cameras 
and the WB camera. We finally normalize them by the median value in the field-of-view.

The seeing during the observations was fairly good although with temporal variations. They were
indeed representative of the typical seeing conditions at the SST that lead to 
scientifically relevant data. However, the training set is slightly suboptimal because of the
limited sampling of seeing conditions. The training set is composed of two spectral scans of 
a quiet Sun region, observed on 2016-09-19 from 10:03 to 10:04 UT, together with another
spectral scan of a region of flux emergence, observed
on the same day from 09:30 to 10:00. The validation set is a spectral scan from the
first run obtained in different seeing conditions. 

Computing the loss function and the final deconvolved image in extended objects that
fill the FOV requires some form of apodization. The computation of the Fourier 
transforms using the Fast Fourier Transform requires periodic functions and 
apodization is a way to force it. We use a modified Hanning window that keeps the
center of the FOV unaffected and only affects the 7 pixels in each border. The
unaffected FOV is then $74 \times 74$ pixels. As a 
consequence of the apodization, it is preferrable to evaluate the loss function of
Eq. (\ref{eq:loss_filter}) in spatial dimensions \citep{1994A&AS..107..243L}.
Although the effect is not very large, one can take into account only
those points in the FOV, removing the apodized part. This can be easily performed
by applying Parseval's theorem.

\subsection{Deconvolution and PSFs}
Figure \ref{fig:examples_sun} shows four representative cases obtained once the neural network is
properly trained. The first column displays the deconvolved images, both the baseline result 
(labeled as MFBD) and the resulting deconvolved image using the PSFs estimated by the neural 
network (labeled as Neural). We show results for the WB and only one of the two NB channels. The two 
NB channels are indeed very similar and their difference is proportional to the 
polarization, which is expected to be very small for many of the observed regions. The first and
third column displays four of the seven available raw frames. It is obvious that the multiframe
blind deconvolution produces much better image quality with only seven frames, but our
architecture can deal seamlessly with an arbitrary number of raw frames. The estimated
PSFs are displayed in the second and fourth column, which are shared by the WB and NB channels.

\section{Conclusions and future outlook}
We have presented a general scheme to train a neural multiframe blind deconvolution 
architecture without the use of supervision. The method makes use only of 
observed images, together with information about the telescope
entrance pupil, the angular pixel size in the camera and the wavelength of
the observations. We have shown, with examples
obtained from the NOT with point-like objects and the SST with extended
objects, that the neural deconvolution generalizes correctly to
unseen images. 
The method also provides as output the instantaneous
wavefronts produced by the atmospheric turbulence, irrespectively of the
number of frames used. The method is extremely fast if compared with standard iterative blind deconvolution methods.
The code for training or evaluation, with the parameters of the networks, is freely 
available\footnote{\texttt{https://github.com/aasensio/unsupervisedMFBD}}.

Given the fundamental ambiguities of inferring wavefronts from
PSFs \citep{paxman19}, we do not consider that our method is especially competitive with
more classical approaches based on, for instance, Shack-Hartmann sensors.
However, our estimation of the instantaneous PSF is adequate for the
deconvolution of images or spectra \citep{2017A&A...608A..76V}.

Our aim in this paper is to present the formalism for the unsupervised 
training. However, we point out that there are several possible ways of improving this work. The first one is
to train an architecture that can blindly deconvolve images from a variety
of telescopes and/or wavelengths. Observations of these telescopes and/or 
wavelengths are needed for the training, though. The formalism remains the same
except on the construction of the OTF from the generalized pupil. In this case, 
one needs to take into account the specific aperture of the telescope and 
the influence of the wavelength on the diffraction limit of the telescope.
Apart from that, we anticipate that conditioning the entrance of the RNNs
feature vector with the telescope properties and the wavelength should be enough.
This can be easily done by concatenating this information on the input vector.

The second potential improvement is to add more training examples that have a larger variety
of objects, especially extending it to other wavelengths of interest. 
However, we note that the convolutional part of the architecture that we have trained is in charge of
obtaining relevant latent features from the image which can produce a good estimation of the
wavefront coefficients. As such, this CNN needs to learn how to be
agnostic to the specific object, which can be difficult if the variability of objects
is increased. This can be easily solved if more information is provided to the neural network. An obvious one is to use a
phase-diversity channel. The combination of the two images contains (theoretically) enough information
to restrict the wavefront. Very preliminary experiments show that this addition strongly 
constrains the problem and produces wavefront coefficients of much better quality,
which might then be competitive with those obtained with classical wavefront sensing methods.
%(Nigul et al., in prep).

Another restriction of our approach is that the input images are currently limited
to be of a fixed size such that $W$ is multiple of 8. This is a consequence of the presence of 
the fully connected GRU and FC networks. This can be potentially solved
by transforming our architecture into a fully convolutional one. Some
convolutional couterparts of RNNs, like the ConvLSTM \citep{convlstm15}, can be used. As well,
the FC network can then be transformed into a fully convolutional network. All networks can be trained with images of
a certain size and, once trained, can be applied to images of any other 
size. For instance, if the input images are of size 128$\times$128, the input 
to the ConvLSTM will have size 16$\times$16, so that at the output we would predict the 
wavefront in 16$\times$16 patches of 8$\times$8 pixels. For computing the loss function one would need 
a way to deal with these spatially variant PSFs. One option would be to compute
the loss function locally in each patch and add them together.

Finally, although the GRU behaves correctly in our case, its serial character makes it
slightly slow to train and cannot be run in paralell. Recurrent neural networks have been 
overcome in recent years by the use of more robusts approaches. We plan to study the application 
of Transformers \citep{transformers17} based on the idea of neural attention to this problem, which 
can better exploit the time information of the observations.

\begin{acknowledgements}
We thank Michiel van Noort for several invaluable discussions during the 
development of this work and for insisting on the interest of
a machine learning approach to infer wavefronts in addition to the deconvolved 
image. We thank \'Alex Oscoz, 
Roberto L\'opez and Jorge Andr\'es Prieto for providing the
FastCam@NOT datasets and Jaime de la Cruz Rodr\'{\i}guez for providing the CRISP@SST datasets.
This study was also discussed
in the workshop \emph{Studying magnetic-field-regulated heating in the 
solar chromosphere} (team 399)
at the International Space Science Institute (ISSI) in Switzerland.
This paper is based on observations made with the Nordic Optical
Telescope operated by the Nordic Optical Telescope Scientific 
Association in the Spanish Observatorio del Roque de los Muchachos
of the Instituto de Astrof\'{\i}sica de Canarias. 
We are very grateful to the ING staff and the IAC Support
Astronomers Group for their efforts.
We acknowledge financial support from the Spanish Ministerio de Ciencia, Innovaci\'on y 
Universidades through project PGC2018-102108-B-I00 and FEDER funds.
This research has made use of NASA's Astrophysics Data System Bibliographic Services.
We acknowledge the community effort devoted to the development of the following 
open-source packages that were
used in this work: \texttt{numpy} \citep[\texttt{numpy.org},][]{numpy20}, 
\texttt{matplotlib} \citep[\texttt{matplotlib.org},][]{matplotlib}, 
\texttt{PyTorch} \citep[\texttt{pytorch.org},][]{pytorch19} and \texttt{h5py} (\texttt{h5py.org}).
\end{acknowledgements}

% \bibliographystyle{aa}
% \bibliography{biblio}

\end{document}